\documentclass[a4paper,10pt]{article}

\usepackage[dvips]{graphicx}
\usepackage{epsfig,amsmath,amssymb,verbatim,mathrsfs,array,layout,textcomp,amssymb,latexsym}

\newcommand{\beq}{\begin{eqnarray}}
\newcommand{\eeq}{\end{eqnarray}}
\newcommand{\nn}{\nonumber}

\def\lsim{\mathrel{\rlap{\lower4pt\hbox{\hskip1pt$\sim$}}
     \raise1pt\hbox{$<$}}}         
\def\gsim{\mathrel{\rlap{\lower4pt\hbox{\hskip1pt$\sim$}}
     \raise1pt\hbox{$>$}}}         

\begin{document}

\begin{titlepage}

\vskip1.5cm
\begin{center}
  {\Large \bf Implications of large dimuon CP asymmetry in $B_{d,s}$ decays
    on minimal flavor violation with low $\tan\beta$}\\
\end{center}
\vskip0.2cm

\begin{center}
{\bf Kfir Blum, Yonit Hochberg and Yosef Nir}

\end{center}
\vskip 8pt

\begin{center}
{\it Department of Particle Physics and Astrophysics\\
Weizmann Institute of Science,\\ Rehovot 76100, Israel} \vspace*{0.3cm}

{\tt  kfir.blum,yonit.hochberg,yosef.nir@weizmann.ac.il}
\end{center}

\vglue 0.3truecm

\begin{abstract}
  \vskip 3pt \noindent The D0 collaboration has recently announced
  evidence for a dimuon CP asymmetry in $B_{d,s}$ decays of order one
  percent. If confirmed, this asymmetry requires new physics. We argue
  that for minimally flavor violating (MFV) new physics, and at low
  $\tan\beta=v_u/v_d$, there are only two four-quark operators
  ($Q_{2,3}$) that can provide the required CP violating effect. The
  scale of such new physics must lie below 260 GeV $\sqrt{\tan\beta}$.
  The effect is universal in the $B_s$ and $B_d$ systems, leading to
  $S_{\psi K}\sim\sin2\beta-0.15$ and $S_{\psi\phi}\sim0.25$. The
  effects on $\epsilon_K$ and on electric dipole moments are negligible.
  The most plausible mechanism is tree-level scalar exchange. MFV
  supersymmetry with low $\tan\beta$ will be excluded. Finally, we
  explain how a pattern of deviations from the Standard Model
  predictions for $S_{\psi\phi}$, $S_{\psi K}$ and $\epsilon_K$ can be
  used to test MFV and, if MFV holds, to probe its structure in
  detail.
\end{abstract}

\end{titlepage}

\section{Introduction}
The D0 collaboration has recently announced evidence for new CP
violating physics in semileptonic $B$ decays \cite{Abazov:2010hv}:
\beq\label{eq:aslexp}
(a_{\rm SL}^b)^{D0}&=&(-9.6\pm2.5\pm1.5)\times10^{-3},
\eeq
to be compared with the Standard Model (SM) prediction \cite{Lenz:2006hd}:
\beq\label{eq:aslthe}
(a_{\rm SL}^b)^{\rm SM}&=&(-0.23^{+0.05}_{-0.06})\times10^{-3}.
\eeq
The measured asymmetry is a combination of the asymmetries in $B_d^0$
and $B_s^0$ decays \cite{Abazov:2010hv}:
\beq\label{eq:aslds} a_{\rm SL}^b&=&
(0.51\pm0.04)a_{\rm SL}^d+(0.49\pm0.04)a_{\rm SL}^s.
\eeq
To explain the difference between the experimental result
(\ref{eq:aslexp}) and the SM prediction (\ref{eq:aslthe}), a new
physics contribution to $B_s-\overline{B_s}$ and/or
$B_d-\overline{B_d}$ mixing is required that is comparable in size to
the SM contribution and carries a new phase of order one.

The fact that, so far, no evidence for new physics in $K$, $D$ and
$B_d$ meson decays has been established implies that the flavor
structure of new physics at a scale $\lsim10^3$ TeV is highly
constrained. This situation is suggestive that perhaps such new
physics carries no new sources of flavor violation beyond the Yukawa
matrices of the SM. This idea, which can be formulated in a rigorous
mathematical way \cite{D'Ambrosio:2002ex}, became known as minimal
flavor violation (MFV)
\cite{Hall:1990ac,Chivukula:1987py,Buras:2000dm}.

The MFV hypothesis does not exclude the possibility of new CP
violating phases, beyond the Kobayashi-Maskawa phase of the SM
\cite{Colangelo:2008qp,Ellis:2007kb,Kagan:2009bn}. In models with more
than a single Higgs doublet, and in particular in the large
$\tan\beta$ limit, such that the bottom Yukawa coupling is of order
one, there is a rather large number of such new phases that could be
large. In contrast, in a single Higgs doublet, or even in multi-Higgs
doublet models where $\tan\beta\ll m_t/m_b$, the situation concerning
CP violation is much more constrained. In this work we study the
implications of the experimental measurement (\ref{eq:aslexp}) for the
latter class of models.

The plan of this paper is as follows. In Section \ref{sec:mfv} we
introduce the effective four-quark operators of interest, and the
flavor suppression factors that accompany them when minimal flavor
violation is imposed. In Section \ref{sec:low} we focus our attention
on MFV models with $\tan\beta\ll m_t/m_b$, find the operators that can
account for a large dimuon CP asymmetry and obtain the resulting
predictions. In Sections \ref{sec:eps} and \ref{sec:edm} we show that
the effects of the relevant operators on, respectively, CP violation
in $K^0-\overline{K^0}$ mixing and electric dipole moments are
negligible. In Section \ref{sec:sus} we argue that a large dimuon CP
asymmetry would exclude the MFV class of the supersymmetric Standard
Model at low $\tan\beta$. In Section \ref{sec:pat} we explain how the
pattern of CP violation in neutral $K$, $B_d$ and $B_s$ meson mixing
can be used to test MFV or to probe its detailed structure. We
conclude in Section \ref{sec:con}.

\section{Minimal Flavor Violation}
\label{sec:mfv}
The effects of new physics at a high energy scale ($\Lambda\gg m_W$)
on $B_q-\overline{B_q}$ mixing can be studied in an effective operator
language. A complete set of four quark operators relevant to
$B_s-\overline{B_s}$ transitions is given by
\beq\label{eq:effope}
Q_1^{sb}&=&\bar b^\alpha_L\gamma_\mu s^\alpha_L \bar
b^\beta_L\gamma_\mu s^\beta_L,\ \ \
\widetilde{Q}_1^{sb}=\bar b^\alpha_R\gamma_\mu s^\alpha_R \bar
b^\beta_R\gamma_\mu s^\beta_R,\nn\\
Q_2^{sb}&=&\bar b^\alpha_R s^\alpha_L \bar
b^\beta_R s^\beta_L,\ \ \
\widetilde{Q}_2^{sb}=\bar b^\alpha_L s^\alpha_R \bar
b^\beta_L s^\beta_R,\nn\\
Q_3^{sb}&=&\bar b^\alpha_R s^\beta_L \bar
b^\beta_R s^\alpha_L,\ \ \
\widetilde{Q}_3^{sb}=\bar b^\alpha_L s^\beta_R \bar
b^\beta_L s^\alpha_R,\nn\\
Q_4^{sb}&=&\bar b^\alpha_R s^\alpha_L \bar
b^\beta_L s^\beta_R,\ \ \
Q_5^{sb}=\bar b^\alpha_R s^\beta_L \bar
b^\beta_L s^\alpha_R.
\eeq
Here $d_L(d_R)$ represent $SU(2)$-doublets (singlets), and
$\alpha,\beta$ are color-indices. The effective Hamiltonian is given
by
\beq
{\cal H}_{\rm eff}^{\Delta B=\Delta S=2}
=\frac{1}{\Lambda^2}\left(\sum_{i=1}^5 z_i Q_i
+\sum_{i=1}^3\tilde z_i \widetilde Q_i\right).
\eeq
For the new physics to give a contribution to the mixing amplitude
that is of order $0.22$ of the SM one [see Eq. (\ref{eq:hsfit})], we
need that at least one of the following conditions will be satisfied:
\beq\label{eq:npeqsm}
|z_1|&\sim&1.2\times 10^{-5} \left(\frac{\Lambda}{{\rm TeV}}\right)^2,\nn\\
|z_2|&\sim&5.5\times 10^{-6} \left(\frac{\Lambda}{{\rm TeV}}\right)^2,\nn\\
|z_3|&\sim&2.0\times 10^{-5} \left(\frac{\Lambda}{{\rm TeV}}\right)^2,\nn\\
|z_4|&\sim&2.0\times 10^{-6} \left(\frac{\Lambda}{{\rm TeV}}\right)^2,\nn\\
|z_5|&\sim&5.3\times 10^{-6} \left(\frac{\Lambda}{{\rm TeV}}\right)^2,
\eeq
(or a value of $|\tilde z_i|$ similar to the one given for the
corresponding $|z_i|$.) We thus learn that (\ref{eq:aslexp}) gives an
upper bound on the scale of the relevant new physics:
\beq\label{eq:upperlambda}
\Lambda\lsim700\ {\rm TeV}.
\eeq

We now impose the MFV principle. Since we are interested in $B_q$
mesons, we work in the down mass basis:
\beq\label{eq:yuyd}
Y_d={\rm diag}(y_d,y_s,y_b),\ \ \ Y_u=V^\dagger\times{\rm
  diag}(y_u,y_c,y_t),
\eeq
where $V$ is the CKM matrix. We further define
\beq\label{eq:auad}
A_d\equiv Y_d Y_d^\dagger\approx{\rm diag}(0,0,y_b^2),\ \ \ A_u\equiv
Y_uY_u^\dagger=V^\dagger\times{\rm
  diag}(0,0,y_t^2)\times V.
\eeq
We now write the MFV form of the $z_i$ coefficients. For each $z_i$,
we include the leading term that does give new CP violation, for
$\tan\beta<m_t/m_b$:
\beq\label{eq:zimfv}
z_1&=&r_1^+ (A_u)_{32}^2 + r_1^-(A_u)_{32}[A_u,A_d]_{32},\nn\\
z_{2,3}&=&r_{2,3}(v^2/\Lambda^2)(Y_d^\dagger A_u)_{32}^2,\nn\\
z_{4,5}&=&r_{4,5}^+ (Y_d^\dagger A_u)_{32}(A_u Y_d)_{32}
+r_{4,5}^- (Y_d^\dagger[A_u,A_d])_{32}(A_u Y_d)_{32}.
\eeq
The $\tilde z_i$ coefficients are highly suppressed. We introduce a
$v^2/\Lambda^2$ factor into $z_{2,3}$ to take into account the fact
that these two operators break $SU(2)_L$ as a triplet. The
coefficients $r_{1,4,5}^+$ are real.

Inserting the expressions (\ref{eq:yuyd}) and (\ref{eq:auad}) into
Eq. (\ref{eq:zimfv}), we obtain
\beq\label{eq:zmfv}
\frac{z_1}{y_t^4(V_{ts}V_{tb}^*)^2}&=&r_1^+ - r_1^- y_b^2,\nn\\
\frac{z_{2,3}}{y_t^4(V_{ts}V_{tb}^*)^2}&=&r_{2,3}(v^2/\Lambda^2)y_b^2,\nn\\
\frac{z_{4,5}}{y_t^4(V_{ts}V_{tb}^*)^2}&=&r_{4,5}^+ y_by_s
-r_{4,5}^- y_b^3y_s.
\eeq

We make the following conclusions:
\begin{enumerate}
\item MFV implies that the $z_i$ coefficients are suppressed by, at
least, $(V_{ts}V_{tb}^*)^2\sim 0.002$. Consequently, (\ref{eq:aslexp})
gives an upper bound on the scale of MFV new physics:
\beq\label{eq:upperlambmfv}
\Lambda_{\rm MFV}\lsim30\ {\rm TeV}.
\eeq
\item If $\tan\beta$ is not very large, there is further suppression
  of the CP violating contributions by a factor $y_b^2$, leading to
\beq\label{eq:upperlamcpv}
\Lambda_{\rm MFV}\lsim500\ {\rm GeV}\ \tan\beta.
\eeq
\item If $r_i^+\gsim r_i^-$, then the phase provided by any of
  $Q_{1,4,5}$ is suppressed by at least $y_b^2$.
\item If {\it all} $r_i$ are of the same order, then the leading
  contribution comes from $Q_1$ and (for $\tan\beta$ not large) is
  approximately CP conserving.
\end{enumerate}

\section{Small $\tan\beta$}
\label{sec:low}
We now focus our attention on $\tan\beta={\cal O}(1)$. Note that
$\tan\beta=1$ in all single Higgs doublet models.

Barring the possibility that $r_i^+\ll r_i^-$ which is, first,
difficult to realize and, second, implies that $r_i^-\ll1$, so that
the new physics contribution suffers further suppression, the only
operators that can give a large CP violating effect in
$B_s-\overline{B_s}$ mixing are $Q_{2,3}$.  As is evident in Eq.
\eqref{eq:npeqsm}, the contribution of $Q_3$ at the low scale is
suppressed in comparison to that of $Q_2$.  Since these operators
share the same MFV structure, Eq. \eqref{eq:zmfv}, we focus our
attention henceforth on $Q_2$.

For $Q_2$ to give a dominant contribution in the MFV case, $z_1$ must
be highly suppressed. This is the case when, for example, the new
physics contribution comes from scalar exchange. As concerns the
competition between $Q_2$ and $Q_4$, the first will dominate if either
$r_4\ll r_2$ or
\beq\label{eq:twofour}
2.7\,y_s/y_b\lsim v^2/\Lambda^2\ \ \ \Longrightarrow\ \ \ \Lambda\lsim 1\
{\rm TeV}
\eeq
where we took into account the relative RGE enhancement of $Q_4$. Given
the upper bound of Eq.  (\ref{eq:upperlamcpv}), and taking into
account that both $z_2$ and $z_4$ are suppressed more strongly than
the ${\cal O}[y_t^4 y_b^2 (V_{tb}^*V_{ts})^2]$ suppression that leads
to Eq. (\ref{eq:upperlamcpv}) (the first by $v^2/\Lambda^2$ and the
latter by $y_s/y_b$), we learn that in much of the regime where either
$Q_2$ or $Q_4$ contribute to $B_s-\overline{B_s}$ mixing comparably to
the SM, we expect $Q_2$ to be comparable to or even dominate over
$Q_4$. The condition then for an ${\cal O}(0.22)$ CP violating
contribution is
\beq\label{eq:qtwo}
\Lambda_{Q_2}\lsim260\ {\rm GeV}\ \sqrt{\tan\beta}.
\eeq

The $Q_2$ dominance, which is necessary to explain (\ref{eq:aslexp})
with low $\tan\beta$ MFV physics, has further interesting
consequences. In particular, it implies that the new physics
contributions are the same in the $B_d$ and $B_s$ systems
\cite{Kagan:2009bn}. More explicitly, defining
\beq\label{eq:defhsigma}
M_{12}^{d,s}=(M_{12}^{d,s})^{\rm SM}(1+h_{d,s}e^{2i\sigma_{d,s}}),
\eeq
the $Q_2$ dominance predicts
\beq\label{eq:univ}
h_b\equiv h_d=h_s,\ \ \ \sigma_b\equiv\sigma_d=\sigma_s.
\eeq
The viability of such a scenario, taking into account all relevant
data about $B_d-\overline{B_d}$ and $B_s-\overline{B_s}$ mixing, was
investigated in Ref. \cite{Ligeti:2010ia}. It was found that it can be
accommodated by the data, even though it is not the most preferred
scenario experimentally. Fitting the data with the assumption
(\ref{eq:univ}), the preferred range of parameters is
\cite{Ligeti:2010ia}
\beq\label{eq:hsfit}
h_b\approx0.22\pm0.06,\ \ \ \sigma_b\approx2.3\pm0.3.
\eeq
It is useful also to rewrite Eq. (\ref{eq:hsfit}) as
\beq
r_b^2 e^{2i\theta_b}\equiv 1+h_b
e^{2i\sigma_b}\approx(1.0\pm0.1)\times e^{-(0.22\pm0.06)i}.
\eeq

This is a rather predictive scenario. Its most significant predictions
can be derived from the following relations:
\beq\label{eq:univmod}
\Delta m_q&=&r_b^2\Delta m_q^{\rm SM},\nn\\
\Delta \Gamma_q&=&\Delta\Gamma_q^{\rm SM}\cos2\theta_b,\nn\\
a_{\rm SL}^q&=&{\cal I}m\left[\Gamma_{12}^q/\left(M_{12}^{q,{\rm SM}}
    r_b^2 e^{2i\theta_b} \right)\right]
\approx-\left(\Gamma_{12}^q/M_{12}^q\right)^{\rm
  SM}(\sin2\theta_b/r_b^2),\nn\\
S_{\psi K}&=&\sin\left(2\beta+2\theta_b\right),\nn\\
S_{\psi\phi}&=&\sin\left(2\beta_s-2\theta_b\right).
\eeq
We obtain:
\beq
\Delta m_q&\approx&(1.0\pm0.1)\Delta m_q^{\rm SM},\nn\\
\Delta \Gamma_q&\approx&(0.98^{+0.01}_{-0.02})\Delta\Gamma_q^{\rm SM},\nn\\
a_{\rm
  SL}^q&\approx&(0.22\pm0.07)\left(\Gamma_{12}^q/M_{12}^q\right)^{\rm
  SM},\nn\\
S_{\psi K}&\approx&0.65\pm0.05,\nn\\
S_{\psi\phi}&\approx&0.25\pm0.06.
\eeq
For the estimate of $S_{\psi K}$, we used the CKMfitter result
\cite{ckmfitter}, $\sin2\beta\approx0.80\pm0.03$. The fact that, for
$h_d=h_s$ and $\sigma_d=\sigma_s$, a negative shift in $S_{\psi K}$ is
correlated with a positive shift in $S_{\psi\phi}$, was first pointed
out in Ref. \cite{Buras:2008nn}.

Note that by assuming that the new physics affects only $M_{12}^q$,
the $B_s$-related observables in Eq. (\ref{eq:univmod}) automatically
satisfy (neglecting $\beta_s$) the relation \cite{Grossman:2009mn}
\beq
\frac{a_{\rm SL}^s}{S_{\psi\phi}/(1-S_{\psi\phi}^2)^{1/2}}=
  -\frac{|\Delta\Gamma_s|}{\Delta m_s}.
\eeq
%

\section{CP violation in $K^0-\overline{K}^0$ mixing}
\label{sec:eps}
Within the MFV framework, the existence of the four-quark term,
\beq
\frac{z_2^{bs}}{\Lambda^2}
\bar b_R^\alpha s_L^\alpha\bar b_R^\beta s_L^\beta,
\eeq
which contributes to CP violation in $B_s-\overline{B_s}$ mixing,
requires the existence of another four-quark term,
\beq
\frac{z_2^{sd}}{\Lambda^2}
\bar s_R^\alpha d_L^\alpha\bar s_R^\beta d_L^\beta,
\eeq
which contributes to CP violation in $K^0-\overline{K}^0$ mixing. The
MFV principle relates $z_2^{sd}$ to $z_2^{bs}$:
\beq
\frac{z_2^{sd}}{z_2^{bs}}=\frac{y_s^2}{y_b^2}
\frac{(V_{td}V_{ts}^*)^2}{(V_{ts}V_{tb}^*)^2}.
\eeq
Thus, the ratio between the imaginary parts of the new physics
contributions is given by
\beq\label{eq:imbsimsd}
\frac{{\cal I}m(z_2^{sd})}{{\cal I}m(z_2^{bs})}=\frac{y_s^2}{y_b^2}
\left|\frac{V_{td}}{V_{tb}}\right|^2
\frac{\sin(2\sigma_b-2\beta)}{\sin(2\sigma_b)}
  \approx1.4\times10^{-8}.
\eeq

For the $z_2^{bs}$ term to provide $h_b$ and $\sigma_b$ of Eq.
(\ref{eq:hsfit}), we need ${\cal
  I}m(z_2^{bs})\sim5.5\times10^{-6}(\Lambda/{\rm TeV})^2$, see Eq.
(\ref{eq:npeqsm}). Together with Eq.  (\ref{eq:imbsimsd}), we obtain
\beq\label{eq:imzsd}
\frac{{\cal I}m(z_2^{sd})}{\Lambda^2}\sim8\times10^{-14}\ {\rm
  TeV}^{-2}.
\eeq
This prediction should be compared to the range allowed by the
$\epsilon_K$ constraint \cite{Bona:2007vi},
\beq
\frac{{\cal I}m(z_2^{sd})}{\Lambda^2}\in\left[-5.1,+9.3\right]
\times10^{-11}\ {\rm TeV}^{-2},
\eeq
which corresponds to a maximal new physics contribution to
$\epsilon_K$ of order ten percent. We conclude that in our framework,
of $Q_2$-dominated CP violation from MFV new physics, the effect on
$\epsilon_K$ is of order a permill and therefore
negligible.

\section{Electric Dipole Moments}
\label{sec:edm}
In this section we investigate whether experimental bounds on electric
dipole moments (EDMs) constrain the four-fermi operators that are
relevant to our study. We use the results of Ref. \cite{Ellis:2008zy}.
A related recent discussion is given in Ref. \cite{Batell:2010qw}
which, however, focuses on MFV models where the $B_s$ effects are
enhanced over those of $B_d$.

Let us consider four-Fermi operators that play a role in EDMs:
\beq
{\cal L}_{4f} = \sum_{i,j} C_{ij}(\bar d_i d_i)(\bar d_j i \gamma_5
d_j),
\eeq
where the coefficients $C_{ij}$ are imaginary and of mass dimension
$-2$. Their contributions to the Mercury EDM are given by
\beq\label{eq:dHg}
d_{H_g}\supset -1.4\cdot 10^{-5} e\ {\rm GeV^2} \left[0.5\frac{
    C_{11}}{m_d}+3.3 \kappa \frac{C_{21}}{m_s}
  +(1-0.25 \kappa)\frac{C_{31}}{m_b}\right],
\eeq
where $\kappa = 0.5 \pm 0.25$. The experimental bound
\cite{Griffith:2009zz},
\beq\label{eq:dHgbound}
\left|d_{H_g}\right|< 3.1 \times 10^{-29}\ e\ {\rm cm},
\eeq
when imposed on each term in \eqref{eq:dHg} independently, gives
\beq
C_{11} &\lsim& 10^{-6}\ {\rm TeV}^{-2},\nn\\
C_{21} &\lsim& 10^{-5}\ {\rm TeV}^{-2},\nn\\
C_{31} &\lsim& 10^{-3}\ {\rm TeV}^{-2}. \eeq
We now focus on the $Q_2$ operator in the MFV framework. It is
convenient to define
\beq
T_{ij}= V_{ti}^* V_{tj} y_t^2 y_{d_j}.
\eeq
Then, in the basis (\ref{eq:yuyd}), we have
\beq
\frac{{\cal I}m z_2}{\Lambda^2}=\frac{v^2}{\Lambda^4}{\cal
  I}m(r_2)(T_{23})^2.
\eeq
On the other hand, in the MFV framework,
\beq
C_{ij}=\frac{{\cal I}m z_2}{\Lambda^2}\,\frac{T_{ii}T_{jj}}{(T_{23})^2}.
\eeq
The EDM bounds then constrain the $Q_2$ contribution (the strongest
constraint comes from $C_{31}$):
\beq \frac{{\rm Im} z_2}{\Lambda^2}\lsim \frac{10}{{\rm TeV}^{2}}. \eeq
Comparing to Eq. (\ref{eq:npeqsm}), we conclude that the EDM bounds
are far from making a significant constraint on the $Q_2$ contribution
to CP violation in $B_s$ mixing. Conversely, in our framework,
of $Q_2$-dominated CP violation from MFV new physics, the effect on
EDMs is about six orders of magnitude below present bounds.

\section{MFV Supersymmetry}
\label{sec:sus}
Within supersymmetry, one can obtain $z_2/\Lambda^2$ in terms of the
masses and mixing angles in the squark sector. Alternatively, one can
use the average squark mass $\tilde m$ and squark mass-insertions
$(\delta^d_{MN})_{23}$, though this is an approximation that is not
very precise when the squark spectrum is far from degeneracy.

As argued above, for MFV and low $\tan\beta$, the only operators that
can give a large enough CP violating effect in $B_s$ mixing are $Q_{2,3}$. In
general, the leading contributions come from gluino-mediated box
diagrams, giving
\beq\label{eq:SUSYz}
\Lambda &=& \tilde m,\\
z_2 &=& -\frac{17}{18}\alpha_s^2 x f_6(x)(\delta_{RL}^d)_{23}^2,\nn\\
z_3 &=& +\frac16 \alpha_s^2 x f_6(x)(\delta_{RL}^d)_{23}^2,\nn\\
\eeq
where $m_{\tilde g}$ is the gluino mass, $x=m_{\tilde g}^2/\tilde
m^2$, and $f_6$ is a known kinematic function, given by
\beq
f_6(x) = \frac{6(1+3x)\ln x +x^3-9x^2-9x+17}{6(x-1)^5}.
\eeq
The expressions for $\tilde z_2$ and $\tilde z_3$ are obtained from
Eq. \eqref{eq:SUSYz} by replacing the indices $RL \rightarrow LR$. To
have $h_b\sim0.22$, we need
\beq\label{eq:delwant}
(\delta_{RL,LR}^d)_{23} \gsim 0.11\left(\frac{\tilde m}{{\rm
      TeV}}\right).
\eeq

Such a large contribution is excluded by the constraints from
$b\rightarrow s\gamma$ and $b\to s\ell^+\ell^-$. The gluino-mediated
$(\delta_{LR}^d)_{23}$-related contributions (for an explicit
expression, see {\it e.g.} Ref. \cite{Gabbiani:1996hi}) provide an
upper bound which, for $x=1$, reads
\cite{Isidori:2010kg,Silvestrini:2007yf}
\beq
|(\delta^d_{LR})_{23}| \lsim0.01 \left(\frac{\tilde m}{{\rm
      TeV}}\right).
\eeq
We conclude that MFV-supersymmetry with $\tan\beta\ll m_t/m_b$ cannot
explain a CP asymmetry in semileptonic $B_s$ decays of order a
percent.

It is actually well known that in general the supersymmetric Standard Model
cannot contribute significantly to $B_s-\overline{B_s}$ mixing via the
$Q_{2,3}$ operators (see {\it e.g.} Ref. \cite{Altmannshofer:2009ne}).
This statement is independent of whether the supersymmetric model is
MFV or not or whether $\tan\beta$ is large or not. What is novel in
our discussion is that we show that in MFV and with low $\tan\beta$,
the $Q_{2,3}$ operators are the {\it only} potential source of a large
effect.

\section{Probing MFV with the pattern of CP violation}
\label{sec:pat}
Our results, when combined with those of Refs.
\cite{Ligeti:2010ia,Buras:2010mh}, show that there is a surprising
variety of scenarios within the MFV framework. Conversely, one can use
the pattern of experimental results to test whether MFV holds and, if
it does, which are the leading operators. (An investigation similar in
spirit to ours, but in the framework of MFV {\it without}
flavor-diagonal CP violating phases, can be found in Ref.
\cite{Buras:2009pj}.)

Our starting point is the flavor suppression factors of the various
$z_i$, presented in Eq. (\ref{eq:zmfv}). In Table~\ref{tab:zmfv} we
rewrite the flavor factors that appear in the ratios between ${\cal
  I}m(z_i^{bd})$, relevant to $S_{\psi K}$, ${\cal I}m(z_i^{sd})$,
relevant to $\epsilon_K$, and ${\cal I}m(z_i^{bs})$, relevant to
$S_{\psi\phi}$. Note that for ${\cal I}m z_i^{bq}$ the relevant term
is the one carrying a new phase, while for ${\cal I}m z_i^{sd}$ the
relevant term can have no new phase. The reason is that a new physics
contribution to $B_s-\overline{B_s}$ [$B_d-\overline{B_d}$] mixing
that is aligned in phase with $(V_{tb}^*V_{ts})^2$
[$(V_{tb}^*V_{td})^2$] leaves $S_{\psi\phi}$ [$S_{\psi K}$] unchanged,
but a new physics contribution to $K^0-\overline{K^0}$ mixing that is
aligned in phase with $(V_{ts}^*V_{td})^2$ does change $\epsilon_K$.
In what follows we assume that the flavor-less coefficients
$r_{1,4,5}^\pm$ and $r_{2,3}$ are of the same order, and do not
affect the suppression pattern depicted in Table \ref{tab:zmfv}.

\begin{table}[t]
  \caption{The magnitude of the flavor factors that appear in the $z_i$ coefficients
    which are relevant to CP violating observables.}
\label{tab:zmfv}
\begin{center}
\begin{tabular}{c|cccc} \hline\hline
  \rule{0pt}{1.2em}%
  $i$ & ${\cal I}m z_i^{bs}$ & ${{\cal I}m z_i^{bd}}/{{\cal I}m z_i^{bs}}$ &
  ${{\cal I}m z_i^{sd}}/{{\cal I}m z_i^{bs}}$ \cr \hline
1 & $y_b^2 y_t^4|V_{tb}V_{ts}|^2$ & $|V_{td}/V_{ts}|^2$ &  $(1/y_b)^2|V_{td}/V_{tb}|^2$ \cr
2,3 & $y_b^2 y_t^4 |V_{tb}V_{ts}|^2$ & $|V_{td}/V_{ts}|^2$ &  $(y_s/y_b)^2|V_{td}/V_{tb}|^2$ \cr
4,5 & $y_b^3y_s y_t^4 |V_{tb}V_{ts}|^2$ &
  $(y_d/y_s)|V_{td}/V_{ts}|^2$ & $(y_d/y_b^3)|V_{td}/V_{tb}|^2$  \cr
\hline\hline
\end{tabular}
\end{center}
\end{table}

By examining Table \ref{tab:zmfv} it should be immediately clear that
the pattern of deviations from the SM in $S_{\psi\phi}$, $S_{\psi K}$
and $\epsilon_K$ can tell us whether a new MFV contribution is
dominated by $Q_1$, $Q_{2,3}$ or $Q_{4,5}$. The different patterns are
presented in Table \ref{tab:pattern}, for large $y_b\sim1$ (very large
$\tan\beta$) and $y_b<1$ (low $\tan\beta$). For each case we saturate
the most restrictive observable (``large''), and estimate whether the
effect on the other observables is similarly significant (``large'')
or smaller than present sensitivity (``small''),

\begin{table}[t]
  \caption{The size of new MFV effects on CP violating observables.}
\label{tab:pattern}
\begin{center}
\begin{tabular}{c|ccc|ccc} \hline\hline
  \rule{0pt}{1.2em}%
  & \multicolumn{3}{c|}{$y_b\sim1$} & \multicolumn{3}{c}{$y_b\ll1$} \cr
  $i$ & $S_{\psi\phi}$ & $S_{\psi K}$ & $\epsilon_K$ & $S_{\psi\phi}$
  & $S_{\psi K}$ & $\epsilon_K$  \cr \hline
1 & small & small & large & small & small & large  \cr
2,3 & large & large & small & large & large & small  \cr
4,5 & large & small & large & small & small & large   \cr
\hline\hline
\end{tabular}
\end{center}
\end{table}

Note that we restrict ourselves here to scenarios where at least one
of the systems gives a convincing signal of new physics. A-priori,
there are seven different patterns among the three systems. We find
that three of these -- the two where there is a large effect on
$S_{\psi K}$ simultaneously with a negligible effect on $S_{\psi\phi}$
(with large or small effect on $\epsilon_K$) \cite{Kagan:2009bn} and
the one with large effects in all three systems cannot be realized in
our framework. The other four scenarios can, and each of them directs
us to specific scenarios: Large effect in $\epsilon_K$ and small
effects in $S_{\psi\phi},S_{\psi K}$ is possible with $Q_1$-dominance
or with $Q_{4,5}$-dominance at low $\tan\beta$. Small effect in
$\epsilon_K$ and large effects in $S_{\psi\phi},S_{\psi K}$ is
possible with $Q_{2,3}$-dominance. Large effects in $\epsilon_K$ and
$S_{\psi\phi}$ with a small effect in $S_{\psi K}$ is possible with
$Q_{4,5}$-dominance at very large $\tan\beta$.

\section{Conclusions}
\label{sec:con}
The D0 collaboration has found evidence for CP violation in
semileptonic $B_s$ decays at the level of one percent. If confirmed,
this result requires new, CP and flavor violating physics
\cite{Laplace:2002ik,Beneke:2003az,Ciuchini:2003ww}.

The measurement further provides interesting upper bounds on the scale
of new physics. In general, it must be well below $10^3$~TeV. If the
new physics is minimally flavor violating (MFV), then its scale must
be below about 30~TeV. In MFV single Higgs doublet models the bound is
further strengthened to 500~GeV.

Within multi-Higgs doublet models with $\tan\beta\sim m_t/m_b$, there
is a variety of ways to generate a large CP asymmetry in the
semileptonic $B_s$ decays from flavor-diagonal phases
\cite{Ligeti:2010ia,Buras:2010mh}.  The situation is, however, much
more constrained if $y_b\ll1$. In this case, the only way to generate
a phase of order one is via the $Q_{2,3}$ operators defined in Eq.
(\ref{eq:effope}). The upper bound on the scale is then still
stronger, $\Lambda\lsim260$ GeV $\sqrt{\tan\beta}$. Furthermore, the
effects in the neutral $B_q$ meson mixing are universal in the $d-s$
flavor space. While the effects on the CP conserving observables
$\Delta m_q$ and $\Delta\Gamma_q$ are small, the effects on the CP
violating observables are large. It is interesting to note that a CP
asymmetry in semileptonic $B_s$ decays that is negative and of order
one percent implies that the CP asymmetry in $B\to\psi K_S$ is shifted
down by about $0.15$ from $\sin2\beta$, which is consistent with
present data.

If the new physics contribution appears at the loop level, we expect a
further suppression of $z_2$ by a factor of $1/16\pi^2$. Upper bounds
on the scale of new physics, such as the one in Eq.
(\ref{eq:upperlamcpv}), become stronger by a factor of $4\pi$, a
dangerously low scale. There is also a need to avoid a situation where
$r_1^+\gsim r_2$, because in this case, the new physics phase,
$\sigma_b$, will be small. Taking these two points together, we are
led to conclude that to explain $a_{\rm SL}^s\sim0.01$ within MFV
models with $y_b\ll 1$, the most likely mechanism is that of
tree-level exchange of a scalar. Indeed, such models have been
suggested in Refs. \cite{Dobrescu:2010rh,Buras:2010mh} to explain the
D0 result. Both works study, however, models with very large
$\tan\beta$.

Finally, if the evidence from the D0 measurement is confirmed, all MFV
versions of the supersymmetric Standard Model with $\tan\beta\ll
m_t/m_b$ will be excluded.

Independent of whether the D0 measurement is confirmed, we demonstrate
how the measurements of $\epsilon_K$, $S_{\psi K}$ and $S_{\psi\phi}$
can be used to test the MFV hypothesis. For example, a shift in
$S_{\psi K}$ from the SM value that is much larger than $S_{\psi\phi}$
will exclude MFV. If the pattern is consistent with MFV, it can be
used to probe its detailed structure. For example, large effects on
$S_{\psi\phi}$ and on $S_{\psi K}$ and no effect on $\epsilon_K$ will
point towards $Q_{2,3}$-dominance, as discussed in this paper, while
observable effects on $S_{\psi\phi}$ and $\epsilon_K$ with only small
effect on $S_{\psi K}$ will point towards $Q_{4,5}$-dominance and
large $\tan\beta$, as discussed in Ref. \cite{Buras:2010mh}.

Since the announcement of the D0 measurement of $a_{\rm SL}^b$, a
number of works interpreting the results have appeared. We mentioned
above the three works that are most closely related to our study,
Refs. \cite{Ligeti:2010ia,Dobrescu:2010rh,Buras:2010mh}, which study
models where the new physics effect is only in the neutral meson
mixing amplitudes and is MFV. The emphasis of all three works is,
however, on large $\tan\beta$. Refs.
\cite{Dighe:2010nj,Bauer:2010dg} assume that the new physics enters
$\Delta B=1$ processes, while we assume that the new physics affects
only the mixing amplitude. Refs.
\cite{Eberhardt:2010bm,Chen:2010wv,Chen:2010aq,Deshpande:2010hy,Parry:2010ce}
study non-MFV models. (For previous works on non-MFV models, see {\it
  e.g.} \cite{Blanke:2006eb,Blanke:2009am,Blanke:2008zb}.)  

{\bf Note added:} When this work was in final stages of writing,
several additional works interpreting the D0 result have appeared
\cite{Jung:2010ik,Choudhury:2010ya,Ko:2010mn,King:2010np,Delaunay:2010dw,Bai:2010kf,Kubo:2010mh}. 
  
\vspace{0.5cm}
\noindent{\Large \bf Acknowledgments}\\
We thank Oram Gedalia, Daniel Grossman, Michele Papucci and Gilad
Perez for helpful discussions.
The work of Y.N. is supported by the Israel Science Foundation (ISF)
under grant No.~377/07, by the German-Israeli foundation for
scientific research and development (GIF), and by the United
States-Israel Binational Science Foundation (BSF), Jerusalem,
Israel.
%



\begin{thebibliography}{99}

\bibitem{Abazov:2010hv}
  V.~M.~Abazov {\it et al.}  [The D0 Collaboration],
  arXiv:1005.2757 [hep-ex].

\bibitem{Lenz:2006hd}
  A.~Lenz and U.~Nierste,
  JHEP {\bf 0706}, 072 (2007)
  [arXiv:hep-ph/0612167].

\bibitem{D'Ambrosio:2002ex}
  G.~D'Ambrosio, G.~F.~Giudice, G.~Isidori and A.~Strumia,
  Nucl.\ Phys.\  B {\bf 645}, 155 (2002)
  [arXiv:hep-ph/0207036].

\bibitem{Hall:1990ac}
  L.~J.~Hall and L.~Randall,
  Phys.\ Rev.\ Lett.\  {\bf 65}, 2939 (1990);

\bibitem{Chivukula:1987py}
  R.~S.~Chivukula and H.~Georgi,
  Phys.\ Lett.\  B {\bf 188}, 99 (1987);

\bibitem{Buras:2000dm}
A.~J.~Buras, P.~Gambino, M.~Gorbahn, S.~Jager and L.~Silvestrini,
  Phys.\ Lett.\  B {\bf 500}, 161 (2001)
  [arXiv:hep-ph/0007085];

\bibitem{Colangelo:2008qp}
  G.~Colangelo, E.~Nikolidakis and C.~Smith,
  Eur.\ Phys.\ J.\  C {\bf 59}, 75 (2009)
  [arXiv:0807.0801 [hep-ph]].

\bibitem{Ellis:2007kb}
  J.~R.~Ellis, J.~S.~Lee and A.~Pilaftsis,
  Phys.\ Rev.\  D {\bf 76}, 115011 (2007)
  [arXiv:0708.2079 [hep-ph]].

\bibitem{Kagan:2009bn}
  A.~L.~Kagan, G.~Perez, T.~Volansky and J.~Zupan,
  Phys.\ Rev.\  D {\bf 80}, 076002 (2009)
  [arXiv:0903.1794 [hep-ph]].

\bibitem{Ligeti:2010ia}
  Z.~Ligeti, M.~Papucci, G.~Perez and J.~Zupan,
  arXiv:1006.0432 [hep-ph].

\bibitem{ckmfitter}
CKMfitter Group (J. Charles et al.),
Eur. Phys. J. C41, 1-131 (2005) [hep-ph/0406184],
updated results and plots available at: http://ckmfitter.in2p3.fr

\bibitem{Buras:2008nn}
  A.~J.~Buras and D.~Guadagnoli,
  Phys.\ Rev.\  D {\bf 78}, 033005 (2008)
  [arXiv:0805.3887 [hep-ph]].

\bibitem{Grossman:2009mn}
  Y.~Grossman, Y.~Nir and G.~Perez,
  Phys.\ Rev.\ Lett.\  {\bf 103}, 071602 (2009)
  [arXiv:0904.0305 [hep-ph]].

\bibitem{Bona:2007vi}
  M.~Bona {\it et al.}  [UTfit Collaboration],
  JHEP {\bf 0803}, 049 (2008)
  [arXiv:0707.0636 [hep-ph]].

\bibitem{Ellis:2008zy}
  J.~R.~Ellis, J.~S.~Lee and A.~Pilaftsis,
  JHEP {\bf 0810}, 049 (2008)
  [arXiv:0808.1819 [hep-ph]].

\bibitem{Batell:2010qw}
  B.~Batell and M.~Pospelov,
  arXiv:1006.2127 [hep-ph].

\bibitem{Griffith:2009zz}
  W.~C.~Griffith, M.~D.~Swallows, T.~H.~Loftus, M.~V.~Romalis, B.~R.~Heckel and E.~N.~Fortson,
  Phys.\ Rev.\ Lett.\  {\bf 102}, 101601 (2009).

\bibitem{Gabbiani:1996hi}
  F.~Gabbiani, E.~Gabrielli, A.~Masiero and L.~Silvestrini,
  Nucl.\ Phys.\  B {\bf 477}, 321 (1996)
  [arXiv:hep-ph/9604387].

\bibitem{Isidori:2010kg}
  G.~Isidori, Y.~Nir and G.~Perez,
  Ann. Rev. Nucl. Part. Sci., in press [arXiv:1002.0900 [hep-ph]].

\bibitem{Silvestrini:2007yf}
  L.~Silvestrini,
  Ann.\ Rev.\ Nucl.\ Part.\ Sci.\  {\bf 57}, 405 (2007)
  [arXiv:0705.1624 [hep-ph]].

\bibitem{Altmannshofer:2009ne}
  W.~Altmannshofer, A.~J.~Buras, S.~Gori, P.~Paradisi and D.~M.~Straub,
  Nucl.\ Phys.\  B {\bf 830}, 17 (2010)
  [arXiv:0909.1333 [hep-ph]].

\bibitem{Buras:2010mh}
  A.~J.~Buras, M.~V.~Carlucci, S.~Gori and G.~Isidori,
  arXiv:1005.5310 [hep-ph].

\bibitem{Buras:2009pj}
  A.~J.~Buras and D.~Guadagnoli,
  Phys.\ Rev.\  D {\bf 79}, 053010 (2009)
  [arXiv:0901.2056 [hep-ph]].

\bibitem{Laplace:2002ik}
  S.~Laplace, Z.~Ligeti, Y.~Nir and G.~Perez,
  Phys.\ Rev.\  D {\bf 65}, 094040 (2002)
  [arXiv:hep-ph/0202010].

\bibitem{Beneke:2003az}
  M.~Beneke, G.~Buchalla, A.~Lenz and U.~Nierste,
  Phys.\ Lett.\  B {\bf 576}, 173 (2003)
  [arXiv:hep-ph/0307344].

\bibitem{Ciuchini:2003ww}
  M.~Ciuchini, E.~Franco, V.~Lubicz, F.~Mescia and C.~Tarantino,
  JHEP {\bf 0308}, 031 (2003)
  [arXiv:hep-ph/0308029].

\bibitem{Dobrescu:2010rh}
  B.~A.~Dobrescu, P.~J.~Fox and A.~Martin,
  arXiv:1005.4238 [hep-ph].

\bibitem{Dighe:2010nj}
  A.~Dighe, A.~Kundu and S.~Nandi,
  arXiv:1005.4051 [hep-ph].

\bibitem{Bauer:2010dg}
  C.~W.~Bauer and N.~D.~Dunn,
  arXiv:1006.1629 [hep-ph].

\bibitem{Eberhardt:2010bm}
  O.~Eberhardt, A.~Lenz and J.~Rohrwild,
  arXiv:1005.3505 [hep-ph].

\bibitem{Chen:2010wv}
  C.~H.~Chen and G.~Faisel,
  arXiv:1005.4582 [hep-ph].

\bibitem{Chen:2010aq}
  C.~H.~Chen, C.~Q.~Geng and W.~Wang,
  arXiv:1006.5216 [hep-ph].

\bibitem{Deshpande:2010hy}
  N.~G.~Deshpande, X.~G.~He and G.~Valencia,
  arXiv:1006.1682 [hep-ph].

\bibitem{Parry:2010ce}
  J.~K.~Parry,
  arXiv:1006.5331 [hep-ph].

\bibitem{Blanke:2006eb}
  M.~Blanke, A.~J.~Buras, A.~Poschenrieder, S.~Recksiegel, C.~Tarantino, S.~Uhlig and A.~Weiler,
  JHEP {\bf 0701}, 066 (2007)
  [arXiv:hep-ph/0610298].

\bibitem{Blanke:2009am}
  M.~Blanke, A.~J.~Buras, B.~Duling, S.~Recksiegel and C.~Tarantino,
  Acta Phys.\ Polon.\  B {\bf 41}, 657 (2010)
  [arXiv:0906.5454 [hep-ph]].

\bibitem{Blanke:2008zb}
  M.~Blanke, A.~J.~Buras, B.~Duling, S.~Gori and A.~Weiler,
  JHEP {\bf 0903}, 001 (2009)
  [arXiv:0809.1073 [hep-ph]].
  
\bibitem{Jung:2010ik}
  M.~Jung, A.~Pich and P.~Tuzon,
  arXiv:1006.0470 [hep-ph].

\bibitem{Choudhury:2010ya}
  D.~Choudhury and D.~K.~Ghosh,
  arXiv:1006.2171 [hep-ph].

\bibitem{Ko:2010mn}
  P.~Ko and J.~h.~Park,
  arXiv:1006.5821 [hep-ph].

\bibitem{King:2010np}
  S.~F.~King,
  arXiv:1006.5895 [hep-ph].

\bibitem{Delaunay:2010dw}
  C.~Delaunay, O.~Gedalia, S.~J.~Lee and G.~Perez,
  arXiv:1007.0243 [hep-ph].

\bibitem{Bai:2010kf}
  Y.~Bai and A.~E.~Nelson,
  arXiv:1007.0596 [hep-ph].

\bibitem{Kubo:2010mh}
  J.~Kubo and A.~Lenz,
  arXiv:1007.0680 [hep-ph].






\end{thebibliography}
\end{document}